\documentclass[preprint,aps,pra,superscriptaddress,showpacs,preprintnumbers]{revtex4-1}
\usepackage[usenames,dvipsnames]{color}
\usepackage{amssymb}
\usepackage{graphicx}
\usepackage{cancel}
\usepackage{amsmath}
\usepackage{braket}
\usepackage{xfrac}
\usepackage{calligra}
\DeclareMathAlphabet{\mathcalligra}{T1}{calligra}{m}{n}
\DeclareFontShape{T1}{calligra}{m}{n}{<->s*[2.2]callig15}{}

\newcommand{\scripty}[1]{\ensuremath{\mathcalligra{#1}}}

\DeclareMathAlphabet{\mathpzc}{OT1}{pzc}{m}{it}

\begin{document}

\title{Bath-induced band decay of a Hubbard lattice gas}
\author{D. Chen}
\author{C. Meldgin}
\author{B. DeMarco}
\affiliation{Department of Physics, University of Illinois, 1110 W Green St., Urbana, IL 61801}
\date{\today}

\begin{abstract}
Dissipation is introduced to a strongly interacting ultracold bosonic gas in the Mott-insulator regime of a 3D spin-dependent optical lattice.  A weakly interacting superfluid comprised of atoms in a state that does not experience the lattice potential acts as a dissipative bath coupled to the lattice atoms via collisions.  Lattice atoms are excited to higher-energy bands via Bragg transitions, and the resulting bath-induced decay is measured using the atomic quasimomentum distribution.  A competing but slower intrinsic decay mechanism arising from collisions between lattice atoms is also investigated.  The measured bath-induced decay rate is compared with the predictions of a weakly interacting model with no free parameters.  The presence of intrinsic decay, which cannot be accommodated within this framework, signals that strong interactions may play a central role in the lattice-atom dynamics.
\end{abstract}


\maketitle

\section{Introduction}

Dissipation plays an essential role in determining the behavior of many quantum systems.  In the form of decoherence, dissipation is deleterious and an obstacle to activities that involve controlling quantum states, such as quantum information processing \cite{Palma08031996}.  Conversely, the effects of dissipation can be advantageous. For example, dissipation can be used to cool many-particle systems into manifestly quantum regimes via coupling to a reservoir. Furthermore, new paradigms have emerged for engineering dissipation to give rise to desired quantum states \cite{diehl2011topology,0295-5075-100-3-30007,diehl2008quantum,PhysRevLett.107.080503} and even as a resource for universal quantum computing \cite{verstraete2009quantum}. Intense research into manipulating dissipation is ongoing, inspired both by these applications and by the many open questions regarding the dynamics of dissipation, especially in strongly interacting systems.

Ultracold gases are remarkably dissipation-free, closed quantum systems and thus an ideal platform for harnessing and studying engineered dissipation.  Here, we introduce dissipation to strongly correlated atoms confined in an optical lattice.  Using a 3D spin-dependent lattice, we engineer a low-entropy, dissipative bath that interacts with a strongly correlated thermal lattice gas prepared in the Mott-insulator regime of the Bose-Hubbard (BH) model.  Previous experiments probing dissipation and entropy exchange in species-dependent potentials have utilized 1D lattices and explored the weakly interacting regime \cite{catani2009entropy,PhysRevLett.111.070401}.

The bath in our measurements affects the lattice gas similarly to the way in which the electromagnetic (EM) vacuum causes decay of excited electronic states in an atom via spontaneous emission. Fluctuations of the electric field of the EM vacuum couple electronic states through the electric dipole interaction, while in our system, collisions with the bath couple lattice atoms to different bands.  Because the EM vacuum is a zero-entropy state, the dipole interaction only causes decay to lower energy states.  Likewise, since the bath in our experiment is a weakly interacting superfluid and thus a low-entropy reservoir, lattice atoms exclusively decay to lower energy bands.  Irreversible decay of electronic states only happens when the spontaneously emitted photon escapes and does not interact with the atom, otherwise coherent vacuum Rabi oscillations occur \cite{PhysRevLett.76.1800}.  We achieve irreversible decay in our system by tuning the trap depth so that the bath atoms involved in inter-band transitions can escape from the gas.

\section{Experimental Procedure}

To carry out these measurements, an unequal mixture of $^{87}$Rb atoms in the ${\big|}F=1,m_F=-1{\big\rangle}$ and $\Ket{1,0}$ hyperfine states are confined in a 1064~nm dipole trap and cooled to $\left(109\pm6\right)$~nK.  At this temperature, the $N^{(l)}=\left(13\pm2\right)\times10^3$ atoms in the $\Ket{1,-1}$ (lattice) state are above the critical temperature for Bose condensation, while the $N^{(b)}=\left(31\pm7\right)\times10^3$ atoms in the $\Ket{1,0}$ (bath) state are below the critical temperature with 30--40\% condensate fraction.  The geometric mean of the dipole trap frequencies is $\left(75\pm6\right)$~Hz and the trap depth is $\left(420\pm60\right)\;k_B\times$nK.  A 10~G magnetic field is applied to the atom gas to suppress spin-changing collisions.

\begin{figure}
\includegraphics[width=1\columnwidth]{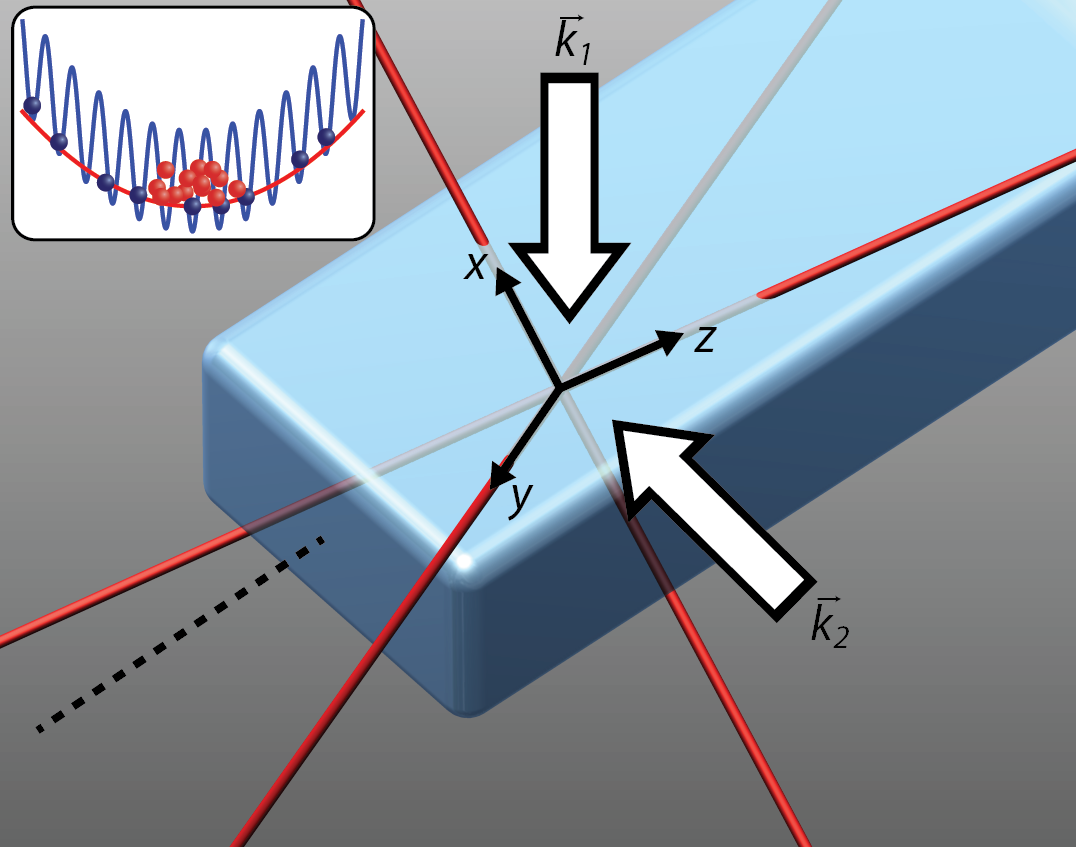}
\caption{The spin-dependent lattice is formed from three pairs of 790~nm laser beams (red lines). None of three lattice directions ($x$, $y$, and $z$) are parallel to the imaging axis (dashed line).  Given the experimental geometry, the $y$ and $z$ directions overlap in the imaging plane.  The Bragg beams propagate along $\vec{k}_1$ and $\vec{k}_2$, so that their wavevector difference lies along the $x$ direction.  The inset shows that the lattice atoms (blue spheres) experience the lattice potential, while the bath atoms (red spheres) are only harmonically trapped by a crossed dipole trap formed from 1064~nm laser beams (not shown).  \label{fig:0}}
\end{figure}

A spin-dependent, cubic optical lattice formed from three pairs of counter-propagating $\lambda=790$~nm laser beams is slowly superimposed on the gas over 50~ms (Fig.~\ref{fig:0}). Detailed information regarding this lattice and our apparatus can be found in Refs.~\cite{PhysRevLett.111.063002,mckay2010thermometry}; similar 1D spin-dependent optical lattices have also been used to study superfluid mixtures \cite{PhysRevLett.105.045303}.  The lattice potential depth is proportional to $\left|m_F\right|$, and thus the lattice atoms experience the lattice potential while the bath atoms are only harmonically trapped, as shown in Fig.~\ref{fig:0}. The lattice atoms realize the strongly interacting BH model with tunneling energy $t$ and interaction energy $U$ controlled by the lattice potential depth \cite{PhysRevLett.81.3108}, and the bath atoms form a spatially overlapping, weakly interacting superfluid bath \cite{PhysRevLett.111.063002}.

We set the lattice potential depth along two lattice directions ($y$ and $z$) to 22.5~$E_R$ and along the third direction ($x$) to $s=\left(13.5\text{--}18\right)$~$E_R$, where $E_R=\hbar^2\pi^2/2 m d$ is the recoil energy, $m$ is the atomic mass, $h=2 \pi \hbar$ is Planck's constant, and $d=\lambda/2$ is the lattice spacing.  At zero temperature, the lattice atoms would be in the Mott insulator phase for the lattice potential depth we explore in this work.  The lattice potential depth is calibrated within a 7\% systematic uncertainty using Kapitza-Dirac diffraction.  The bandgap (between the ground and first excited bands) $E_{\text{bg}}$ for the lattice atoms along the $x$ direction ranges from $\left(1100\text{--}1300\right)$~$k_B\times$nK, which is greater than the dipole trap depth for the bath atoms.  The corresponding bandgap along the $y$ and $z$ directions is 1500~$k_B\times$nK.  The difference in $x$ and $y,z$ bandgaps, which is greater than the excited bandwidths, prevents inter-particle collisions between lattice atoms in excited bands from transferring energy between lattice directions.

For the regime explored in this work, the temperature of the lattice atoms is greater than the ground lattice bandwidth and much less than $E_{\text{bg}}/k_B$, so that they approximately fill the ground band in the lattice but do not occupy excited bands.  We use semiclassical thermodynamics (neglecting interactions) to estimate the temperature of the lattice atoms in the lattice \cite{mckay:2009}.  We assume that the lattice turn-on is isentropic, and we solve self-consistently for the chemical potential and temperature that reproduce the number of lattice atoms and their entropy before turning on the lattice.  We find that the predicted temperature in the lattice ranges from 50 to 130 times the Hubbard tunneling energy $t$ (for the ground band) and is 0.07 times the bandgap for the $x$ direction, given the $s$ sampled in this work.

\section{Band Excitation and Decay}

After turning on the lattice, approximately 40\% of the lattice atoms are transferred to the first excited band along $x$ using Bragg transitions driven by two additional laser beams 500~GHz detuned from the transition to the $5P_{3/2}$ excited electronic state. As shown in Fig.~\ref{fig:0}, the wavevector difference between the Bragg beams is aligned with the $x$-direction of the lattice and has magnitude $1.4\pi/d$.  During the 200~$\mu$s pulse of Bragg light, lattice atoms are transferred to the first excited band along $x$ while leaving them in the ground band along $y$ and $z$.  The frequency difference between the beams is centered on $\left(22\text{--}27\right)$~kHz (corresponding to $E_{\text{bg}}/h$) and swept by 2~kHz during the Bragg pulse.  Using a two-band Rabi model, we calculate that the excitation probability varies from 0.45--0.65 across the band.  We measure that off-resonant and excitation through the first excited band leads to 10\% of the atoms driven to the second excited band.

\begin{figure}
\includegraphics[width=1\columnwidth]{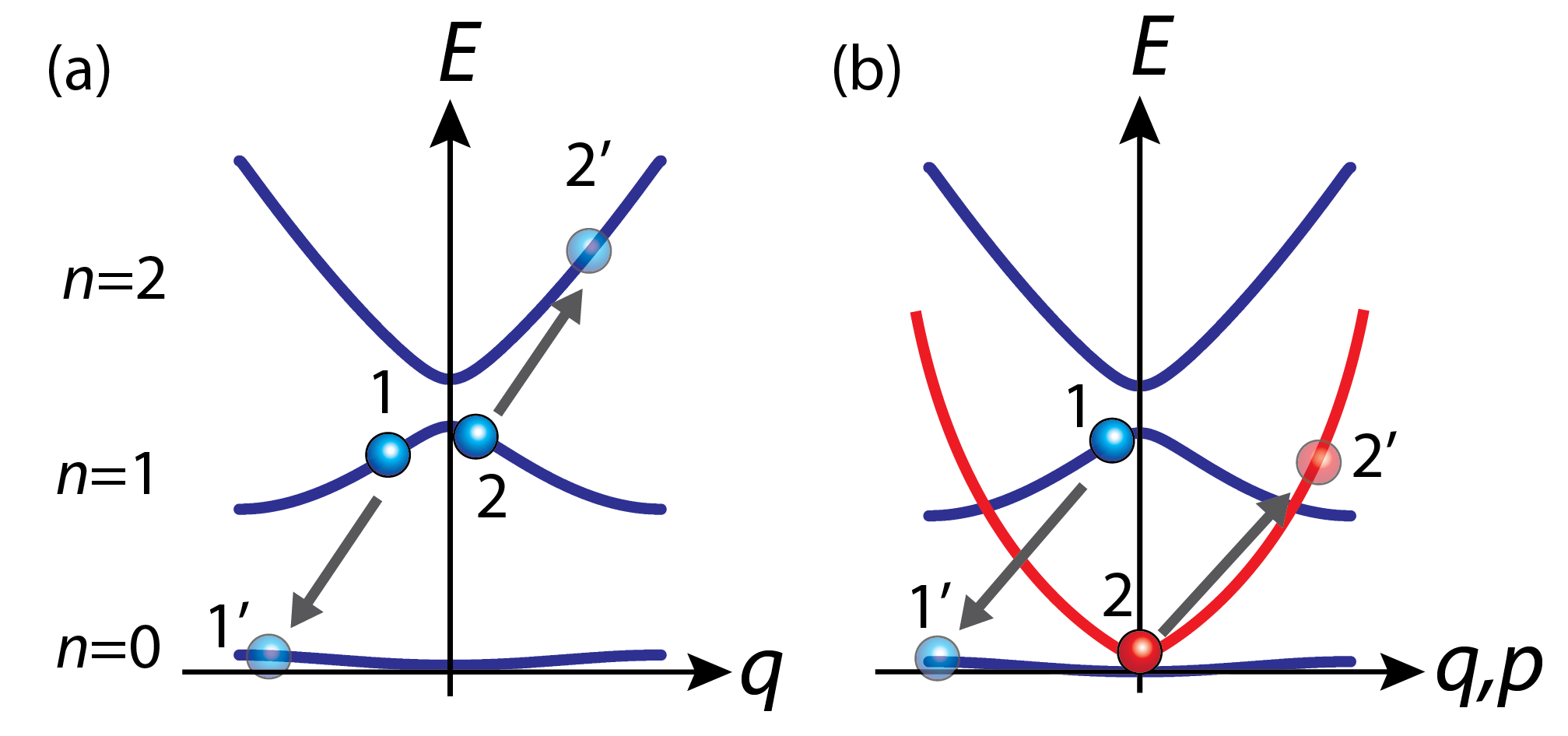}
\caption{Schematic representation of possible band decay processes.  The intrinsic decay arising from collisions between lattice atoms (blue spheres) is shown in (a).  The energy of bands $n$ as a function of quasimomentum $q$ is shown using blue lines for the $x$ direction of the lattice.  Atoms begin in initial states $1$ and $2$ and scatter to states $1'$ and $2'$ through a collision.  The bath-induced decay process that arises through inter-species collisions is shown in (b).  The Bogoliubov dispersion (red lines) as a function of momentum $p$ for the bath atoms (red spheres) is superimposed on the band structure.  The bath atoms scatter into particle-like Bogoliubov excitations that can escape from the trap, which has a depth smaller than $E_{\text{bg}}$. \label{decays}}
\end{figure}

The gas is held in the lattice for time $t_{\text{hold}}$ after the lattice atoms are excited while decay to the ground band occurs.  We previously showed that thermalization between ground-band atoms and the bath is suppressed because conservation of energy and momentum in inter-species collisions is not possible \cite{PhysRevLett.111.063002}.  However, this constraint does not apply to lattice atoms in higher-energy bands.  There are therefore two possible decay channels for excited lattice atoms: collisions between lattice atoms (Fig.~\ref{decays}a) and collisions between lattice and bath atoms (Fig.~\ref{decays}b).

In the intrinsic decay mechanism, two lattice atoms in the first excited ($n=1$) band collide, and one atom decays to the ground band while the other is promoted to the second excited ($n=2$) band.  Intrinsic decay has previously been investigated in very strong 3D \cite{PhysRevLett.99.200405} and in 1D \cite{PhysRevA.87.063638} species-independent optical lattices.  In a non-interacting system, the intrinsic decay channel is suppressed for the relatively high $s$ we employ because of energy and momentum conservation constraints.  As we will discuss, the presence of intrinsic decay may signal the influence of strong inter-particle interactions.  The bath-induced decay channel can be understood as a lattice atom colliding with a Bogoliubov quasiparticle initially at low energy and momentum in the superfluid bath.  After the collision, the quasiparticle is scattered to higher energy and momentum, and the lattice atom decays to the ground ($n=0$) band.  Based on energy conservation, the energy of the excited quasiparticle will approximately be equal to $E_{\text{bg}}$ (within the bandwidth), which is larger than the dipole trap depth for the bath atoms.  Furthermore, for our range of experimental parameters, a quasiparticle with energy equal to $E_{\text{bg}}$ has a momentum more than 6 times greater than $\sqrt{2m\mu}$, where $\mu$ is the chemical potential of the bath.  Therefore, the excited quasiparticle is particle-like, and the bath-induced decay process can result in a bath atom leaving the trap.

The fraction of lattice atoms in the ground band after $t_{\text{hold}}$ is measured from quasimomentum distributions acquired using bandmapping \cite{mckay:2009} and time-of-flight imaging.  Bandmapping is a procedure that transforms quasimomentum to momentum by turning off the lattice quickly (i.e., in 300~$\mu$s) compared with $h/t$ but slowly with respect to $h/E_{\text{bg}}$. After bandmapping, the trap is quickly turned off and the gas is allowed to freely expand for 10~ms.  During this expansion, the spin components are spatially separated by an applied magnetic field gradient and independently imaged.

Characteristic images of the lattice atoms taken for $s=16.2$~$E_R$ are shown in Fig. 3 before the Bragg excitation (Fig. 3a) and after $t_{\text{hold}}$ (Figs. 3b and c).  Images of the lattice component for two situations are displayed---without (Fig. 3b) and with (Fig. 3c) the bath.  We measure the fraction of lattice atoms in the ground band $N_g^{(l)}/N^{(l)}$ by masking off the first Brillouin zone projected onto the imaging plane (dashed lines in Fig. 3).  Atoms excited by the Bragg transition appear outside of the first Brillouin zone.  We measure $N_g^{(l)}/N^{(l)}$ at each $t_{\text{hold}}$ using the ratio of the integrated optical depth (OD) inside the first Brillouin zone to the total OD for lattice images.

\begin{figure}
\includegraphics[width=1\columnwidth]{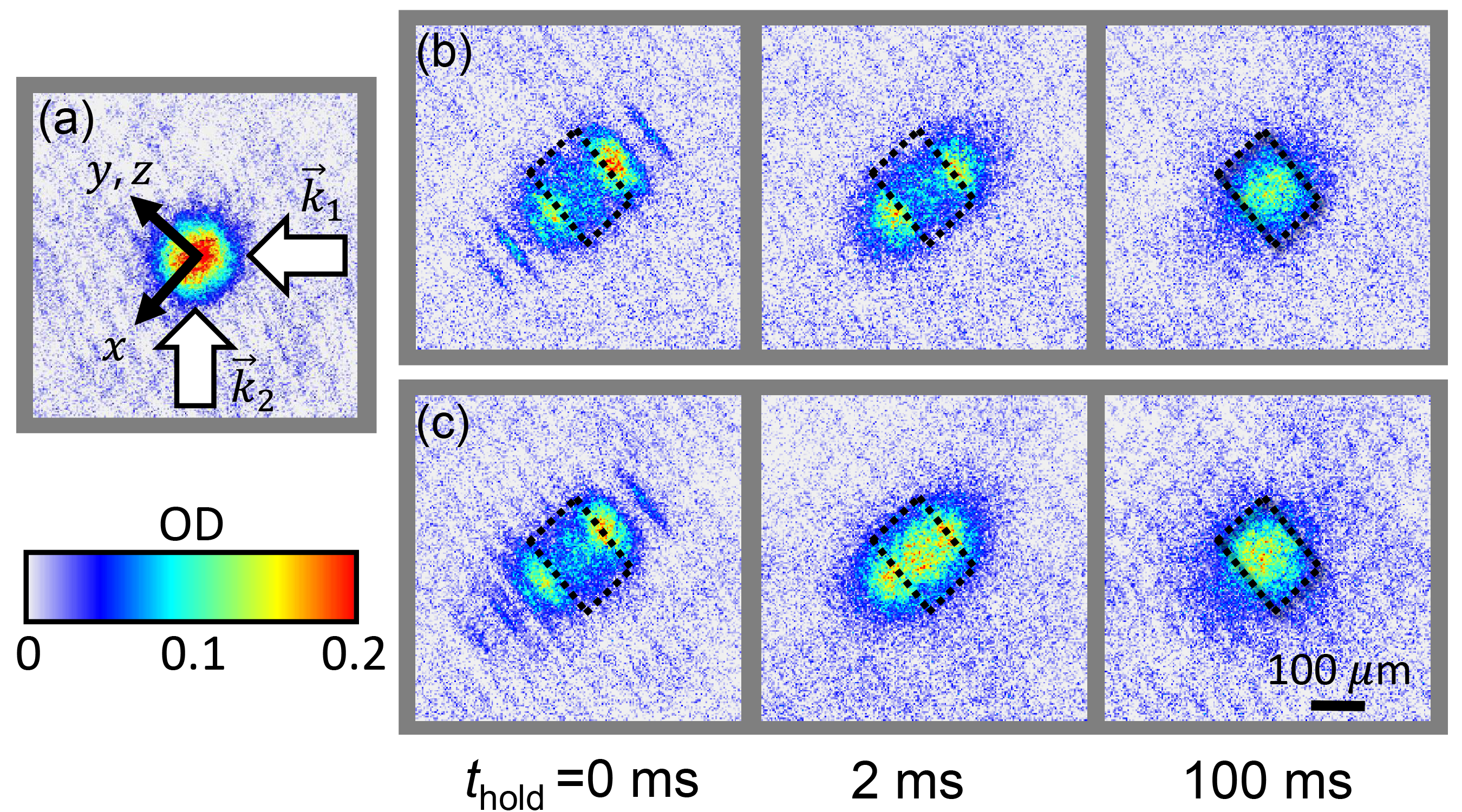}
\caption{The images shown in this figure are of the lattice atoms only, which have been spatially separated from the bath component.  The projection of the geometry of the Bragg laser beams (white arrows) and lattice beams (black arrows) onto the imaging plane is shown superimposed on an image of the lattice atoms before Bragg excitation in (a).  As discussed in the text, the lattice atoms are sufficiently hot so as to uniformly fill the ground band. Images taken without the bath present after waiting $t_{\text{hold}}$ following Bragg excitation are shown in (b), and images taken when the bath is present are shown in (c).  The first Brillouin zone projected onto the imaging plane is displayed using dashed lines.  \label{fig:2}}
\end{figure}

Two systematic issues potentially affect our determination of $N_g^{(l)}/N^{(l)}$.  First, for the imaging direction we use, atoms excited along the $y$ and $z$ directions can appear to be within the Brillouin zone projected onto the imaging plane.  Ideally there are no atoms excited along these directions because the Bragg wavevector is aligned along the $x$ direction and the lattice potential depth configuration prevents exchange of excitation energy between lattice directions.  Second, a systematic error is introduced to measuring $N_g^{(l)}/N^{(l)}$ using our imaging method because bandmapping fails at the band edge and atoms within ground band appear outside the Brillouin zone \cite{mckay:2009}.  Before Bragg excitation, for example, we measure $N_g^{(l)}/N^{(l)}=0.77\pm0.06$, even though the temperature is sufficiently low compared with $E_{\text{bg}}$ such that more than 93\% of the atoms are in the ground band (according to semiclassical thermodynamics).  This error does not significantly affect the measured decay rate since the ground band is homogeneously filled and the excitation is nearly uniform across the Brillouin zone.

The data shown in Fig. 4 for $N_g^{(l)}/N^l$ vs. $t_{\text{hold}}$ at $s=16.2$~$E_R$ demonstrate that the excited lattice atoms decay exponentially after the Bragg excitation, and that the decay rate is enhanced by the bath.  The fraction of atoms in the excited band decays exponentially with and without the bath present because both decay channels involve binary collisions and only one partner in the collisions is measured. The decay time constant is $1/\Gamma_{\text{ll}}$ with the bath absent, where $\Gamma_{\text{ll}}$ is the rate of binary collisions between excited lattice atoms, while the decay time constant with the bath present is $\left(\Gamma_{\text{ll}}+\Gamma_{\text{lb}}\right)^{-1}$, where $\Gamma_{\text{lb}}$ is the rate of binary collisions between bath and excited lattice atoms.  We therefore fit $N_g^{(l)}/N^{(l)}$ to a single exponential decay (lines in Fig. 4) in all cases to determine a decay time constant $\tau$.

\begin{figure}
\includegraphics[width=1\columnwidth]{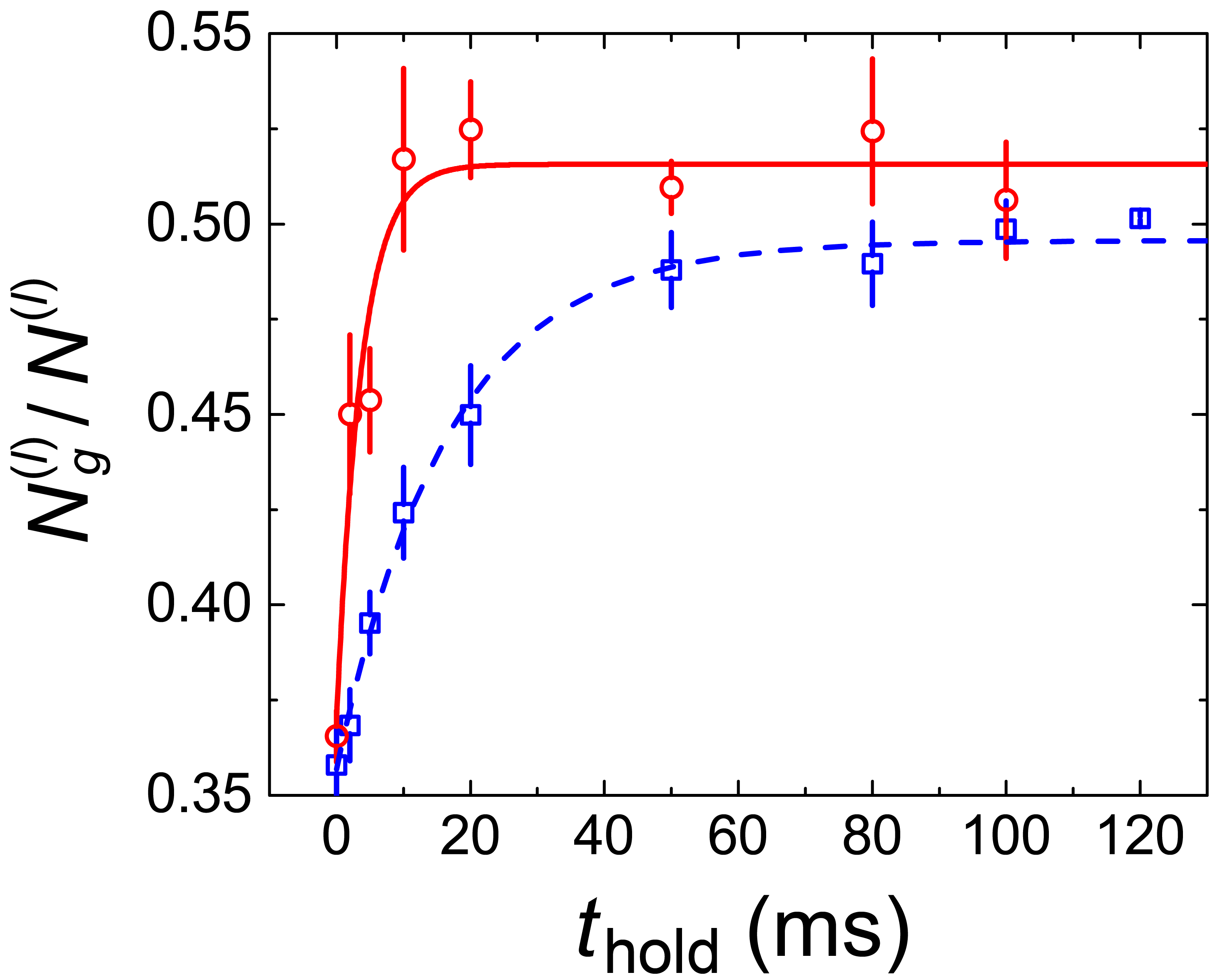}
\caption{Band decay data extracted from images such as those shown in Fig.~3.  Data and corresponding fits are shown for a mixture of lattice and bath states (red circles, solid red line) and only lattice atoms (blue squares, dashed blue line).  The error bars show the standard error in the mean for the 4--6 experimental runs averaged for each point.  \label{fig:3}}
\end{figure}

The fitted decay time constant $\tau$ with and without the bath present is shown in Fig.~5 for $s=13.5$, 16.2, and 18~$E_R$.  For this range of lattice potential depths, the bath-induced decay rate is at least a factor of two larger than the intrinsic decay rate from collisions between lattice atoms.  The measured decay time constants do not vary strongly with $s$.

\begin{figure}
\includegraphics[width=1\columnwidth]{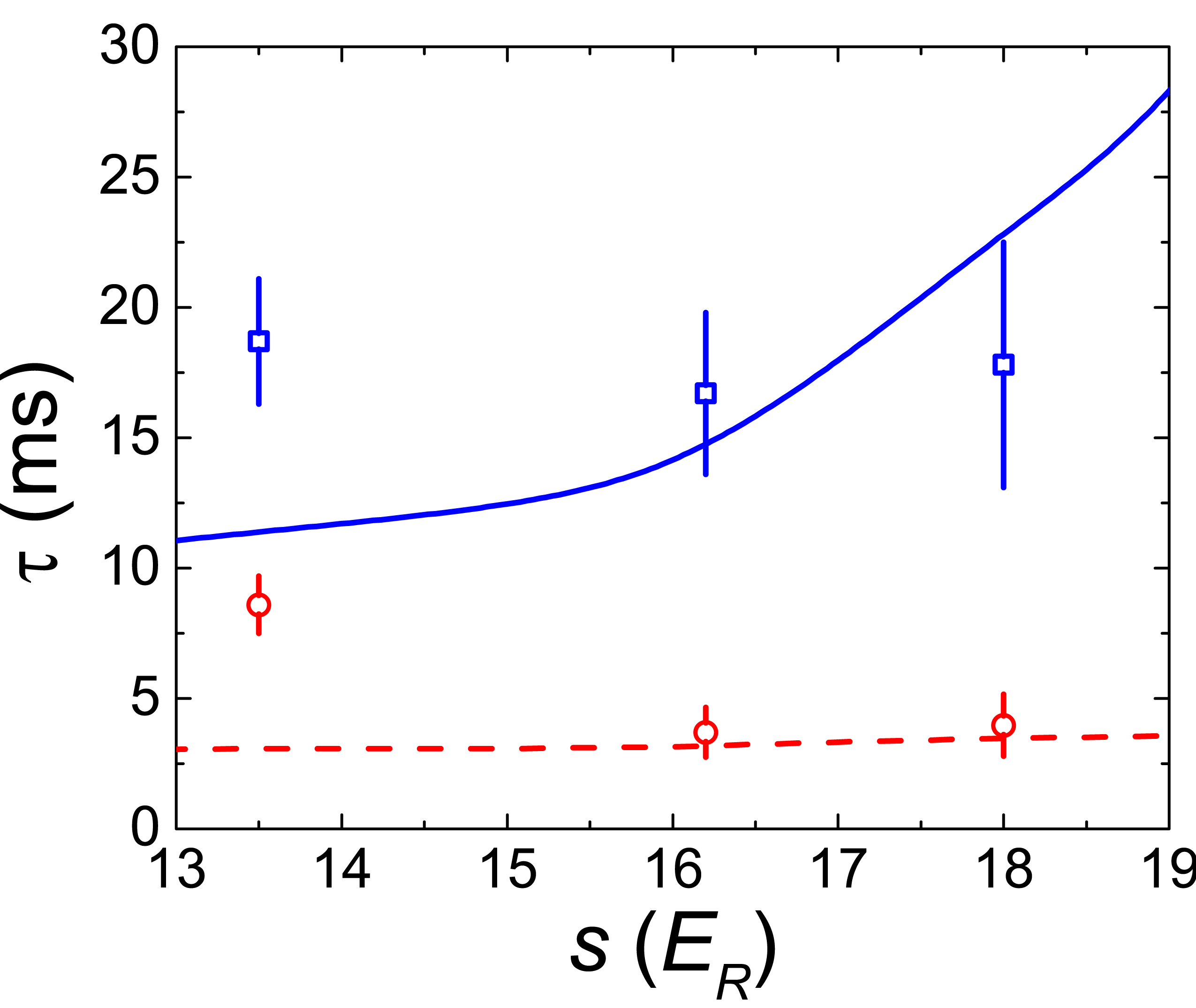}
\caption{Decay time constant measured for lattice atoms only (blue squares) and for a mixture of lattice and bath atoms (red squares) at different lattice potential depths. The error bars are the uncertainty in the fit to data (such as those shown in Fig. 4) used to determine $\tau$.  The solid and dashed lines are $1/\Gamma_{\text{ll}}$ and $\left(\Gamma_{\text{ll}}+\Gamma_{\text{lb}}\right)^{-1}$, which are the FGR predictions.\label{fig:4}}
\end{figure}

We also analyze images of the bath component taken at different $t_{\text{hold}}$.  For these images, a two-component fit is used to measure condensate fraction and $N^{(b)}$.  For all $s$, the condensate fraction for the bath is constant with the experimental noise during the decay process.  Based on the measured change in $N_g^{(l)}/N^{(l)}$, the loss of bath atoms expected during the bath-induced decay is too small to resolve within the noise in $N^{(b)}$ at all $s$.

\section{Fermi's Golden Rule Prediction}

We compare the fitted decay time constants to predictions based on Fermi's golden rule (FGR) in Fig. 5.  We calculate the rate of collisions $\Gamma$ using FGR with the contact potential $V=\left(4\pi a \hbar^2/m \right)\delta^3\left(\vec{r}_i- \vec{r}_j\right)$ for low-energy binary collisions, where $\Gamma$ is the collision rate that leads to decay into the ground band, $\vec{r}_i$ is the location of particle $i$, and $a$ is the scattering length. FGR is evaluated using the Fourier expansion $V = \bigl(4\pi a \hbar^2/mL^3 \bigr)\sum_{\vec{q}} e^{-i\vec{q}\cdot(\vec{r}_i-\vec{r}_j)}$, and we approximate sums over momentum or quasimomentum as $\sum_{\vec{q}}\rightarrow(L^3/2\pi)\int d^3q$.  Single-particle quantum states are used for the lattice component, and we only consider atoms in the $n=0$ and $n=1$ bands.

For the intrinsic decay, we apply FGR to the collision between two lattice atoms assuming that dynamics are frozen along the $y$ and $z$ directions.  The rate for collisions that scatter a pair of atoms with quasimomentum $q_{1_x}$ and $q_{2_x}$ from the first excited band ($n=1$) into quasimomentum $q'_{1_x}$ and $q'_{2_x}$ in the ground ($n=0$) and second excited ($n=2$) band (Fig.~2a) is
\begin{align}
\Gamma_{\text{ll}}=\,&\frac{2\pi}{\hbar} \sum_{q'_{1_x},q'_{2_x}} \left| \Braket{q_{1_x}^{\prime(0)} q_{2_x}^{\prime(2)} | V\left(\vec{r}_1-\vec{r}_2\right) | q_{1_x}^{(1)} q_{2_x}^{(1)} } \right|^2 \nonumber \\
&\times\delta\left[\epsilon^{(0)}(q'_{1_x})+ \epsilon^{(2)}(q'_{2_x})-\epsilon^{(1)}(q_{2_x})-\epsilon^{(1)}(q_{1_x})  \right].
\label{eq:1}\end{align}
For determining $\Gamma_{\text{ll}}$, we treat the lattice atoms as uniform and contained in a cube of volume $L^3$.  We evaluate $\Gamma_{\text{ll}}$ using wavefunctions $\psi_{q_{i_x}}^{(n)}(\vec{r})=\phi_{q_{i_x}}^{(n)}(x)w^{(0)}(y)w^{(0)}(z)$, where $\phi_{q}^{(n)}(x)$ is the Bloch wavefunction in band $n$ with quasimomentum $q$ along lattice direction $x$ and $w^{(n)}(y)$ and $w^{(n)}(z)$ are the Wannier wavefunctions for band $n$ along lattice directions $y$ and $z$.  The coefficients $c^{(n)}_j(q)$ for the plane-wave expansion $\phi_{q}^{(n)}(x)=e^{iqx}\sum_{j=-\infty}^\infty c^{(n)}_j(q)e^{i2kjx}/\linebreak[1]\sqrt{L}$ and the lattice dispersion $\epsilon^{(n)}(q)$ for band $n$ are determined numerically, with $k=2\pi/\lambda$.  We truncate the sum over plane wave components to $|j|\leq3$, approximate the Wannier wavefunctions as the harmonic-oscillator ground states of the lattice wells. We average $\Gamma_{\text{ll}}$ over $q_{1_x}$ and $q_{2_x}$ assuming a uniform distribution of initial quasimomentum, reflecting the high temperature we employ for the lattice component.

Given a non-interacting $\epsilon(q_x)$ and the range of $s$ we sample, this intrinsic decay process should be prevented by the energy-conserving delta function in $\Gamma_{\text{ll}}$ and thus $\Gamma_{\text{ll}}=0$.  Energy cannot be conserved in collisions that scatter non-interacting atoms between bands because of the band structure and bandgaps for the lattice potential depths we use.  The decay we observe is thus evidence that strong interactions may play a central role in dynamics.  To compare with the measured decay rate, we therefore relax energy conservation and represent the delta function in $\Gamma_{\text{ll}}$ by a Gaussian with root-mean-squared radius $U_{11}=\left(4\pi a \hbar^2/m\right)\int d^3\vec{r} \left|w^{(1)}(x)\right|^4 \left|w^{(0)}(y)\right|^4  \left|w^{(0)}(z)\right|^4$, which is the Hubbard interaction energy between atoms in the first excited band along direction $x$ and in the ground band in directions $y$ and $z$.

To compute the bath-induced decay rate, we treat the lattice atoms as confined to single sites of the lattice, which are approximated as three-dimensional harmonic oscillator potentials (Fig.~2b).  The bath atoms are described as a weakly interacting, zero-temperature superfluid (SF).  We consider collisions between a single lattice atom and the bath that cause the lattice atom to decay from the first to ground vibrational level along $x$ and produce a quasiparticle in the bath with momentum $\vec{p}$, and compute
\begin{equation}
\Gamma_{\text{lb}}=\frac{2\pi}{\hbar}\sum_{\vec{p}} \left| \Braket{\Phi_{\vec{p}},\psi_{000} |\sum_j V(\vec{r}_1-\vec{\scripty{r}}_j) |\Phi_0, \psi_{100}} \right|^2 \delta(\epsilon_p-h \nu_x),
\end{equation}
where $\vec{\scripty{r}}_j$ are the coordinates of the bath atoms, $\vec{r}_1$ is the coordinate of the lattice atom, $\Ket{\Phi_0}$ is the ground state of the SF bath, $\Ket{\Phi_{\vec{p}}}$ is the bath with a Bogoliubov quasiparticle of momentum $\vec{p}$ present, $\Ket{\psi_{n_xn_yn_z}}$ is the harmonic oscillator state with quantum numbers $n_x$, $n_y$, and $n_z$, and $\nu_x$ is the vibrational frequency of the lattice well along $x$. The bath is treated as uniform and contained in a cube of volume $L^3$, while the lattice atoms are modeled using localized Wannier states.  Because $h\nu_x$ (which is $E_{\text{bg}}$ in the deep-lattice limit) is much greater than the chemical potential of the bath, we approximate the superfluid (SF) dispersion as $\epsilon_p\approx p^2/2m$ and $\Braket{ \Phi_{\vec{p}}|\sum_{j}e^{i\vec{q}\cdot\vec{\scripty{r}}_j} |\Phi_0 }\approx\sqrt{n_0L^3}\delta_{\vec{p},\vec{q}}$ for calculating the bath-induced decay rate \cite{PhysRevA.61.063608,timmermans1998superfluidity}. We use the density in the center of the trapped SF estimated according to the Thomas-Fermi approximation for $n_0$. The matrix element $\Braket{\psi_{000} |e^{-i\vec{p}\cdot\vec{r}_1} |\psi_{100} }$ is determined using the results in Ref.~\cite{wineland1997experimental}.  Energy is conserved in $\Gamma_{\text{lb}}$ for the bath-induced decay curve shown in Fig. 5.

In Fig.~5, we show predictions for $\tau$ with and without the bath.  We account for both decay channels when the bath is present.  The computed intrinsic time constant shown in Fig.~5 increases for higher $s$ because the phase space for collisions that conserve momentum and energy within $U_{11}$ shrinks.  In contrast, the bath-induced decay rate is largely independent of $s$ because the bandgap, which determines the density of states of quasi-particle excitations in the bath, depends weakly on the lattice potential depth ($E_{\text{bg}}\propto\sqrt{s}$ for large $s$).  The predicted and measured $\tau$ agree within the experimental uncertainty at $s=16$ and $s=18$~$E_R$, but disagree significantly at $s=13.5$~$E_R$. Ultimately, the FGR calculation we implement is limited and should not be expected to entirely model the decay. Fully including tunneling or implementing approaches such as dynamical density-matrix renormalization group methods (see \cite{PhysRevB.66.045114}, for example) to more accurately treat strong interactions requires significant computational resources and is beyond the scope of this work.  Furthermore, we ignore the nonzero temperature of the bath and the dynamics of atoms excited to the $n=2$ band by Bragg transitions or the intrinsic decay process; including these effects would require unavailable computational resources.

\section{Implications for Bath-Induced Cooling}

The bath-induced decay we observe is a key ingredient in an optical-lattice cooling scheme put forward in Ref.~\cite{griessner:2006} and a proposed method for Bose condensation via dissipation \cite{diehl2008quantum}.  The intrinsic decay process we observe (which was not considered in these proposals) may negatively impact the success of such a cooling scheme.  The deleterious impact of intrinsic decay is evident in Fig.~4.  Even though the presence of the bath marginally increases the ground-band fraction to $0.516 \pm 0.007$ when the decay is complete (relative to $0.496\pm0.006$ in its absence, according to the exponential fit), the final ground-band fraction is greatly reduced compared to the state ($N_g^{(l)}/N^{(l)}=0.77\pm0.06$) before Bragg excitation.  Heating despite the bath-induced decay occurs because a small fraction of intrinsic decay events can deposit a large amount of energy into the lattice gas since $E_{\text{bg}}\gg12t$.  An analysis of this process requires a more sophisticated approach to handling the strong interactions between lattice atoms and is beyond the scope of this work.  This problem could be mitigated by a larger ratio $N^{(b)}/N^{(l)}$ (since the bath-induced decay rate is proportional to the bath density) or by using a Feshbach resonance \cite{RevModPhys.82.1225} to transiently suppress the interactions between lattice atoms.

\section{Conclusions}

In conclusion, we have observed intrinsic and bath-induced band decay in a spin-dependent lattice.  The band-induced decay rate is at least double the intrinsic decay rate resulting from collisions between lattice atoms.  The measured decay time constants were compared to a simple FGR model with no free parameters. Our measurements of the intrinsic decay process may be important to proposals for bath-cooling schemes \cite{griessner:2006} and techniques for preparing ultracold gases in excited bands of optical lattices (see Refs. \cite{PhysRevA.83.013610,PhysRevA.88.033615,PhysRevA.72.053604,PhysRevA.81.023605}, for example).  In the future, the lattice-bath interaction we have demonstrated may be a rich platform for exploring the influence of dissipation on strongly correlated dynamics and on exotic many-particle quantum states, such as excited-orbital superfluids \cite{soltan2011quantum}.

\begin{acknowledgements}
This work was supported by the National Science Foundation, Army Research Office, and the DARPA OLE program.  C. Meldgin acknowledges support from an NSF Graduate Fellowship. We thank Andrew Daley, Erich Mueller, and Vito Scarola for helpful discussions.
\end{acknowledgements}

\bibliography{band_decay}

\begin{thebibliography}{27}%
\makeatletter
\providecommand \@ifxundefined [1]{%
 \@ifx{#1\undefined}
}%
\providecommand \@ifnum [1]{%
 \ifnum #1\expandafter \@firstoftwo
 \else \expandafter \@secondoftwo
 \fi
}%
\providecommand \@ifx [1]{%
 \ifx #1\expandafter \@firstoftwo
 \else \expandafter \@secondoftwo
 \fi
}%
\providecommand \natexlab [1]{#1}%
\providecommand \enquote  [1]{``#1''}%
\providecommand \bibnamefont  [1]{#1}%
\providecommand \bibfnamefont [1]{#1}%
\providecommand \citenamefont [1]{#1}%
\providecommand \href@noop [0]{\@secondoftwo}%
\providecommand \href [0]{\begingroup \@sanitize@url \@href}%
\providecommand \@href[1]{\@@startlink{#1}\@@href}%
\providecommand \@@href[1]{\endgroup#1\@@endlink}%
\providecommand \@sanitize@url [0]{\catcode `\\12\catcode `\$12\catcode
  `\&12\catcode `\#12\catcode `\^12\catcode `\_12\catcode `\%12\relax}%
\providecommand \@@startlink[1]{}%
\providecommand \@@endlink[0]{}%
\providecommand \url  [0]{\begingroup\@sanitize@url \@url }%
\providecommand \@url [1]{\endgroup\@href {#1}{\urlprefix }}%
\providecommand \urlprefix  [0]{URL }%
\providecommand \Eprint [0]{\href }%
\providecommand \doibase [0]{http://dx.doi.org/}%
\providecommand \selectlanguage [0]{\@gobble}%
\providecommand \bibinfo  [0]{\@secondoftwo}%
\providecommand \bibfield  [0]{\@secondoftwo}%
\providecommand \translation [1]{[#1]}%
\providecommand \BibitemOpen [0]{}%
\providecommand \bibitemStop [0]{}%
\providecommand \bibitemNoStop [0]{.\EOS\space}%
\providecommand \EOS [0]{\spacefactor3000\relax}%
\providecommand \BibitemShut  [1]{\csname bibitem#1\endcsname}%
\let\auto@bib@innerbib\@empty
\bibitem [{\citenamefont {Palma}\ \emph {et~al.}(1996)\citenamefont {Palma},
  \citenamefont {Suominen},\ and\ \citenamefont {Ekert}}]{Palma08031996}%
  \BibitemOpen
  \bibfield  {author} {\bibinfo {author} {\bibfnamefont {G.~M.}\ \bibnamefont
  {Palma}}, \bibinfo {author} {\bibfnamefont {K.-A.}\ \bibnamefont {Suominen}},
  \ and\ \bibinfo {author} {\bibfnamefont {A.~K.}\ \bibnamefont {Ekert}},\
  }\href {\doibase 10.1098/rspa.1996.0029} {\bibfield  {journal} {\bibinfo
  {journal} {P. Roy. Soc. A-Math Phy.}\ }\textbf {\bibinfo {volume} {452}},\
  \bibinfo {pages} {567} (\bibinfo {year} {1996})}\BibitemShut {NoStop}%
\bibitem [{\citenamefont {Diehl}\ \emph {et~al.}(2011)\citenamefont {Diehl},
  \citenamefont {Rico}, \citenamefont {Baranov},\ and\ \citenamefont
  {Zoller}}]{diehl2011topology}%
  \BibitemOpen
  \bibfield  {author} {\bibinfo {author} {\bibfnamefont {S.}~\bibnamefont
  {Diehl}}, \bibinfo {author} {\bibfnamefont {E.}~\bibnamefont {Rico}},
  \bibinfo {author} {\bibfnamefont {M.~A.}\ \bibnamefont {Baranov}}, \ and\
  \bibinfo {author} {\bibfnamefont {P.}~\bibnamefont {Zoller}},\ }\href@noop {}
  {\bibfield  {journal} {\bibinfo  {journal} {Nat. Phys.}\ }\textbf {\bibinfo
  {volume} {7}},\ \bibinfo {pages} {971} (\bibinfo {year} {2011})}\BibitemShut
  {NoStop}%
\bibitem [{\citenamefont {Kordas}\ \emph {et~al.}(2012)\citenamefont {Kordas},
  \citenamefont {Wimberger},\ and\ \citenamefont
  {Witthaut}}]{0295-5075-100-3-30007}%
  \BibitemOpen
  \bibfield  {author} {\bibinfo {author} {\bibfnamefont {G.}~\bibnamefont
  {Kordas}}, \bibinfo {author} {\bibfnamefont {S.}~\bibnamefont {Wimberger}}, \
  and\ \bibinfo {author} {\bibfnamefont {D.}~\bibnamefont {Witthaut}},\ }\href
  {http://stacks.iop.org/0295-5075/100/i=3/a=30007} {\bibfield  {journal}
  {\bibinfo  {journal} {Europhys. Lett.}\ }\textbf {\bibinfo {volume} {100}},\
  \bibinfo {pages} {30007} (\bibinfo {year} {2012})}\BibitemShut {NoStop}%
\bibitem [{\citenamefont {Diehl}\ \emph {et~al.}(2008)\citenamefont {Diehl},
  \citenamefont {Micheli}, \citenamefont {Kantian}, \citenamefont {Kraus},
  \citenamefont {B{\"u}chler},\ and\ \citenamefont
  {Zoller}}]{diehl2008quantum}%
  \BibitemOpen
  \bibfield  {author} {\bibinfo {author} {\bibfnamefont {S.}~\bibnamefont
  {Diehl}}, \bibinfo {author} {\bibfnamefont {A.}~\bibnamefont {Micheli}},
  \bibinfo {author} {\bibfnamefont {A.}~\bibnamefont {Kantian}}, \bibinfo
  {author} {\bibfnamefont {B.}~\bibnamefont {Kraus}}, \bibinfo {author}
  {\bibfnamefont {H.}~\bibnamefont {B{\"u}chler}}, \ and\ \bibinfo {author}
  {\bibfnamefont {P.}~\bibnamefont {Zoller}},\ }\href@noop {} {\bibfield
  {journal} {\bibinfo  {journal} {Nat. Phys.}\ }\textbf {\bibinfo {volume}
  {4}},\ \bibinfo {pages} {878} (\bibinfo {year} {2008})}\BibitemShut {NoStop}%
\bibitem [{\citenamefont {Krauter}\ \emph {et~al.}(2011)\citenamefont
  {Krauter}, \citenamefont {Muschik}, \citenamefont {Jensen}, \citenamefont
  {Wasilewski}, \citenamefont {Petersen}, \citenamefont {Cirac},\ and\
  \citenamefont {Polzik}}]{PhysRevLett.107.080503}%
  \BibitemOpen
  \bibfield  {author} {\bibinfo {author} {\bibfnamefont {H.}~\bibnamefont
  {Krauter}}, \bibinfo {author} {\bibfnamefont {C.~A.}\ \bibnamefont
  {Muschik}}, \bibinfo {author} {\bibfnamefont {K.}~\bibnamefont {Jensen}},
  \bibinfo {author} {\bibfnamefont {W.}~\bibnamefont {Wasilewski}}, \bibinfo
  {author} {\bibfnamefont {J.~M.}\ \bibnamefont {Petersen}}, \bibinfo {author}
  {\bibfnamefont {J.~I.}\ \bibnamefont {Cirac}}, \ and\ \bibinfo {author}
  {\bibfnamefont {E.~S.}\ \bibnamefont {Polzik}},\ }\href {\doibase
  10.1103/PhysRevLett.107.080503} {\bibfield  {journal} {\bibinfo  {journal}
  {Phys. Rev. Lett.}\ }\textbf {\bibinfo {volume} {107}},\ \bibinfo {pages}
  {080503} (\bibinfo {year} {2011})}\BibitemShut {NoStop}%
\bibitem [{\citenamefont {Verstraete}\ \emph {et~al.}(2009)\citenamefont
  {Verstraete}, \citenamefont {Wolf},\ and\ \citenamefont
  {Cirac}}]{verstraete2009quantum}%
  \BibitemOpen
  \bibfield  {author} {\bibinfo {author} {\bibfnamefont {F.}~\bibnamefont
  {Verstraete}}, \bibinfo {author} {\bibfnamefont {M.~M.}\ \bibnamefont
  {Wolf}}, \ and\ \bibinfo {author} {\bibfnamefont {J.~I.}\ \bibnamefont
  {Cirac}},\ }\href@noop {} {\bibfield  {journal} {\bibinfo  {journal} {Nat.
  Phys.}\ }\textbf {\bibinfo {volume} {5}},\ \bibinfo {pages} {633} (\bibinfo
  {year} {2009})}\BibitemShut {NoStop}%
\bibitem [{\citenamefont {Catani}\ \emph {et~al.}(2009)\citenamefont {Catani},
  \citenamefont {Barontini}, \citenamefont {Lamporesi}, \citenamefont
  {Rabatti}, \citenamefont {Thalhammer}, \citenamefont {Minardi}, \citenamefont
  {Stringari},\ and\ \citenamefont {Inguscio}}]{catani2009entropy}%
  \BibitemOpen
  \bibfield  {author} {\bibinfo {author} {\bibfnamefont {J.}~\bibnamefont
  {Catani}}, \bibinfo {author} {\bibfnamefont {G.}~\bibnamefont {Barontini}},
  \bibinfo {author} {\bibfnamefont {G.}~\bibnamefont {Lamporesi}}, \bibinfo
  {author} {\bibfnamefont {F.}~\bibnamefont {Rabatti}}, \bibinfo {author}
  {\bibfnamefont {G.}~\bibnamefont {Thalhammer}}, \bibinfo {author}
  {\bibfnamefont {F.}~\bibnamefont {Minardi}}, \bibinfo {author} {\bibfnamefont
  {S.}~\bibnamefont {Stringari}}, \ and\ \bibinfo {author} {\bibfnamefont
  {M.}~\bibnamefont {Inguscio}},\ }\href@noop {} {\bibfield  {journal}
  {\bibinfo  {journal} {Phys. Rev. Lett.}\ }\textbf {\bibinfo {volume} {103}},\
  \bibinfo {pages} {140401} (\bibinfo {year} {2009})}\BibitemShut {NoStop}%
\bibitem [{\citenamefont {Scelle}\ \emph {et~al.}(2013)\citenamefont {Scelle},
  \citenamefont {Rentrop}, \citenamefont {Trautmann}, \citenamefont
  {Schuster},\ and\ \citenamefont {Oberthaler}}]{PhysRevLett.111.070401}%
  \BibitemOpen
  \bibfield  {author} {\bibinfo {author} {\bibfnamefont {R.}~\bibnamefont
  {Scelle}}, \bibinfo {author} {\bibfnamefont {T.}~\bibnamefont {Rentrop}},
  \bibinfo {author} {\bibfnamefont {A.}~\bibnamefont {Trautmann}}, \bibinfo
  {author} {\bibfnamefont {T.}~\bibnamefont {Schuster}}, \ and\ \bibinfo
  {author} {\bibfnamefont {M.~K.}\ \bibnamefont {Oberthaler}},\ }\href
  {\doibase 10.1103/PhysRevLett.111.070401} {\bibfield  {journal} {\bibinfo
  {journal} {Phys. Rev. Lett.}\ }\textbf {\bibinfo {volume} {111}},\ \bibinfo
  {pages} {070401} (\bibinfo {year} {2013})}\BibitemShut {NoStop}%
\bibitem [{\citenamefont {Brune}\ \emph {et~al.}(1996)\citenamefont {Brune},
  \citenamefont {Schmidt-Kaler}, \citenamefont {Maali}, \citenamefont {Dreyer},
  \citenamefont {Hagley}, \citenamefont {Raimond},\ and\ \citenamefont
  {Haroche}}]{PhysRevLett.76.1800}%
  \BibitemOpen
  \bibfield  {author} {\bibinfo {author} {\bibfnamefont {M.}~\bibnamefont
  {Brune}}, \bibinfo {author} {\bibfnamefont {F.}~\bibnamefont
  {Schmidt-Kaler}}, \bibinfo {author} {\bibfnamefont {A.}~\bibnamefont
  {Maali}}, \bibinfo {author} {\bibfnamefont {J.}~\bibnamefont {Dreyer}},
  \bibinfo {author} {\bibfnamefont {E.}~\bibnamefont {Hagley}}, \bibinfo
  {author} {\bibfnamefont {J.~M.}\ \bibnamefont {Raimond}}, \ and\ \bibinfo
  {author} {\bibfnamefont {S.}~\bibnamefont {Haroche}},\ }\href {\doibase
  10.1103/PhysRevLett.76.1800} {\bibfield  {journal} {\bibinfo  {journal}
  {Phys. Rev. Lett.}\ }\textbf {\bibinfo {volume} {76}},\ \bibinfo {pages}
  {1800} (\bibinfo {year} {1996})}\BibitemShut {NoStop}%
\bibitem [{\citenamefont {McKay}\ \emph {et~al.}(2013)\citenamefont {McKay},
  \citenamefont {Meldgin}, \citenamefont {Chen},\ and\ \citenamefont
  {DeMarco}}]{PhysRevLett.111.063002}%
  \BibitemOpen
  \bibfield  {author} {\bibinfo {author} {\bibfnamefont {D.~C.}\ \bibnamefont
  {McKay}}, \bibinfo {author} {\bibfnamefont {C.}~\bibnamefont {Meldgin}},
  \bibinfo {author} {\bibfnamefont {D.}~\bibnamefont {Chen}}, \ and\ \bibinfo
  {author} {\bibfnamefont {B.}~\bibnamefont {DeMarco}},\ }\href {\doibase
  10.1103/PhysRevLett.111.063002} {\bibfield  {journal} {\bibinfo  {journal}
  {Phys. Rev. Lett.}\ }\textbf {\bibinfo {volume} {111}},\ \bibinfo {pages}
  {063002} (\bibinfo {year} {2013})}\BibitemShut {NoStop}%
\bibitem [{\citenamefont {McKay}\ and\ \citenamefont
  {DeMarco}(2010)}]{mckay2010thermometry}%
  \BibitemOpen
  \bibfield  {author} {\bibinfo {author} {\bibfnamefont {D.}~\bibnamefont
  {McKay}}\ and\ \bibinfo {author} {\bibfnamefont {B.}~\bibnamefont
  {DeMarco}},\ }\href@noop {} {\bibfield  {journal} {\bibinfo  {journal} {New
  J. Phys.}\ }\textbf {\bibinfo {volume} {12}},\ \bibinfo {pages} {055013}
  (\bibinfo {year} {2010})}\BibitemShut {NoStop}%
\bibitem [{\citenamefont {Gadway}\ \emph {et~al.}(2010)\citenamefont {Gadway},
  \citenamefont {Pertot}, \citenamefont {Reimann},\ and\ \citenamefont
  {Schneble}}]{PhysRevLett.105.045303}%
  \BibitemOpen
  \bibfield  {author} {\bibinfo {author} {\bibfnamefont {B.}~\bibnamefont
  {Gadway}}, \bibinfo {author} {\bibfnamefont {D.}~\bibnamefont {Pertot}},
  \bibinfo {author} {\bibfnamefont {R.}~\bibnamefont {Reimann}}, \ and\
  \bibinfo {author} {\bibfnamefont {D.}~\bibnamefont {Schneble}},\ }\href
  {\doibase 10.1103/PhysRevLett.105.045303} {\bibfield  {journal} {\bibinfo
  {journal} {Phys. Rev. Lett.}\ }\textbf {\bibinfo {volume} {105}},\ \bibinfo
  {pages} {045303} (\bibinfo {year} {2010})}\BibitemShut {NoStop}%
\bibitem [{\citenamefont {Jaksch}\ \emph {et~al.}(1998)\citenamefont {Jaksch},
  \citenamefont {Bruder}, \citenamefont {Cirac}, \citenamefont {Gardiner},\
  and\ \citenamefont {Zoller}}]{PhysRevLett.81.3108}%
  \BibitemOpen
  \bibfield  {author} {\bibinfo {author} {\bibfnamefont {D.}~\bibnamefont
  {Jaksch}}, \bibinfo {author} {\bibfnamefont {C.}~\bibnamefont {Bruder}},
  \bibinfo {author} {\bibfnamefont {J.~I.}\ \bibnamefont {Cirac}}, \bibinfo
  {author} {\bibfnamefont {C.~W.}\ \bibnamefont {Gardiner}}, \ and\ \bibinfo
  {author} {\bibfnamefont {P.}~\bibnamefont {Zoller}},\ }\href {\doibase
  10.1103/PhysRevLett.81.3108} {\bibfield  {journal} {\bibinfo  {journal}
  {Phys. Rev. Lett.}\ }\textbf {\bibinfo {volume} {81}},\ \bibinfo {pages}
  {3108} (\bibinfo {year} {1998})}\BibitemShut {NoStop}%
\bibitem [{\citenamefont {McKay}\ \emph {et~al.}(2009)\citenamefont {McKay},
  \citenamefont {White},\ and\ \citenamefont {DeMarco}}]{mckay:2009}%
  \BibitemOpen
  \bibfield  {author} {\bibinfo {author} {\bibfnamefont {D.}~\bibnamefont
  {McKay}}, \bibinfo {author} {\bibfnamefont {M.}~\bibnamefont {White}}, \ and\
  \bibinfo {author} {\bibfnamefont {B.}~\bibnamefont {DeMarco}},\ }\href@noop
  {} {\bibfield  {journal} {\bibinfo  {journal} {Phys. Rev. A}\ }\textbf
  {\bibinfo {volume} {79}},\ \bibinfo {pages} {063605} (\bibinfo {year}
  {2009})}\BibitemShut {NoStop}%
\bibitem [{\citenamefont {M\"uller}\ \emph {et~al.}(2007)\citenamefont
  {M\"uller}, \citenamefont {F\"olling}, \citenamefont {Widera},\ and\
  \citenamefont {Bloch}}]{PhysRevLett.99.200405}%
  \BibitemOpen
  \bibfield  {author} {\bibinfo {author} {\bibfnamefont {T.}~\bibnamefont
  {M\"uller}}, \bibinfo {author} {\bibfnamefont {S.}~\bibnamefont {F\"olling}},
  \bibinfo {author} {\bibfnamefont {A.}~\bibnamefont {Widera}}, \ and\ \bibinfo
  {author} {\bibfnamefont {I.}~\bibnamefont {Bloch}},\ }\href {\doibase
  10.1103/PhysRevLett.99.200405} {\bibfield  {journal} {\bibinfo  {journal}
  {Phys. Rev. Lett.}\ }\textbf {\bibinfo {volume} {99}},\ \bibinfo {pages}
  {200405} (\bibinfo {year} {2007})}\BibitemShut {NoStop}%
\bibitem [{\citenamefont {Zhai}\ \emph {et~al.}(2013)\citenamefont {Zhai},
  \citenamefont {Yue}, \citenamefont {Wu}, \citenamefont {Chen}, \citenamefont
  {Zhang},\ and\ \citenamefont {Zhou}}]{PhysRevA.87.063638}%
  \BibitemOpen
  \bibfield  {author} {\bibinfo {author} {\bibfnamefont {Y.}~\bibnamefont
  {Zhai}}, \bibinfo {author} {\bibfnamefont {X.}~\bibnamefont {Yue}}, \bibinfo
  {author} {\bibfnamefont {Y.}~\bibnamefont {Wu}}, \bibinfo {author}
  {\bibfnamefont {X.}~\bibnamefont {Chen}}, \bibinfo {author} {\bibfnamefont
  {P.}~\bibnamefont {Zhang}}, \ and\ \bibinfo {author} {\bibfnamefont
  {X.}~\bibnamefont {Zhou}},\ }\href {\doibase 10.1103/PhysRevA.87.063638}
  {\bibfield  {journal} {\bibinfo  {journal} {Phys. Rev. A}\ }\textbf {\bibinfo
  {volume} {87}},\ \bibinfo {pages} {063638} (\bibinfo {year}
  {2013})}\BibitemShut {NoStop}%
\bibitem [{\citenamefont {Zambelli}\ \emph {et~al.}(2000)\citenamefont
  {Zambelli}, \citenamefont {Pitaevskii}, \citenamefont {Stamper-Kurn},\ and\
  \citenamefont {Stringari}}]{PhysRevA.61.063608}%
  \BibitemOpen
  \bibfield  {author} {\bibinfo {author} {\bibfnamefont {F.}~\bibnamefont
  {Zambelli}}, \bibinfo {author} {\bibfnamefont {L.}~\bibnamefont
  {Pitaevskii}}, \bibinfo {author} {\bibfnamefont {D.~M.}\ \bibnamefont
  {Stamper-Kurn}}, \ and\ \bibinfo {author} {\bibfnamefont {S.}~\bibnamefont
  {Stringari}},\ }\href {\doibase 10.1103/PhysRevA.61.063608} {\bibfield
  {journal} {\bibinfo  {journal} {Phys. Rev. A}\ }\textbf {\bibinfo {volume}
  {61}},\ \bibinfo {pages} {063608} (\bibinfo {year} {2000})}\BibitemShut
  {NoStop}%
\bibitem [{\citenamefont {Timmermans}\ and\ \citenamefont
  {C{\^o}t{\'e}}(1998)}]{timmermans1998superfluidity}%
  \BibitemOpen
  \bibfield  {author} {\bibinfo {author} {\bibfnamefont {E.}~\bibnamefont
  {Timmermans}}\ and\ \bibinfo {author} {\bibfnamefont {R.}~\bibnamefont
  {C{\^o}t{\'e}}},\ }\href@noop {} {\bibfield  {journal} {\bibinfo  {journal}
  {Phys. Rev. Lett.}\ }\textbf {\bibinfo {volume} {80}},\ \bibinfo {pages}
  {3419} (\bibinfo {year} {1998})}\BibitemShut {NoStop}%
\bibitem [{\citenamefont {Wineland}\ \emph {et~al.}(1998)\citenamefont
  {Wineland}, \citenamefont {Monroe}, \citenamefont {Itano}, \citenamefont
  {Leibfried}, \citenamefont {King},\ and\ \citenamefont
  {Meekhof}}]{wineland1997experimental}%
  \BibitemOpen
  \bibfield  {author} {\bibinfo {author} {\bibfnamefont {D.~J.}\ \bibnamefont
  {Wineland}}, \bibinfo {author} {\bibfnamefont {C.}~\bibnamefont {Monroe}},
  \bibinfo {author} {\bibfnamefont {W.}~\bibnamefont {Itano}}, \bibinfo
  {author} {\bibfnamefont {D.}~\bibnamefont {Leibfried}}, \bibinfo {author}
  {\bibfnamefont {B.}~\bibnamefont {King}}, \ and\ \bibinfo {author}
  {\bibfnamefont {D.}~\bibnamefont {Meekhof}},\ }\href@noop {} {\bibfield
  {journal} {\bibinfo  {journal} {J. Res. Natl. Inst. Stand. Tech.}\ }\textbf
  {\bibinfo {volume} {103}},\ \bibinfo {pages} {259} (\bibinfo {year}
  {1998})}\BibitemShut {NoStop}%
\bibitem [{\citenamefont {Jeckelmann}(2002)}]{PhysRevB.66.045114}%
  \BibitemOpen
  \bibfield  {author} {\bibinfo {author} {\bibfnamefont {E.}~\bibnamefont
  {Jeckelmann}},\ }\href {\doibase 10.1103/PhysRevB.66.045114} {\bibfield
  {journal} {\bibinfo  {journal} {Phys. Rev. B}\ }\textbf {\bibinfo {volume}
  {66}},\ \bibinfo {pages} {045114} (\bibinfo {year} {2002})}\BibitemShut
  {NoStop}%
\bibitem [{\citenamefont {Griessner}\ \emph {et~al.}(2006)\citenamefont
  {Griessner}, \citenamefont {Daley}, \citenamefont {Clark}, \citenamefont
  {Jaksch},\ and\ \citenamefont {Zoller}}]{griessner:2006}%
  \BibitemOpen
  \bibfield  {author} {\bibinfo {author} {\bibfnamefont {A.}~\bibnamefont
  {Griessner}}, \bibinfo {author} {\bibfnamefont {A.~J.}\ \bibnamefont
  {Daley}}, \bibinfo {author} {\bibfnamefont {S.~R.}\ \bibnamefont {Clark}},
  \bibinfo {author} {\bibfnamefont {D.}~\bibnamefont {Jaksch}}, \ and\ \bibinfo
  {author} {\bibfnamefont {P.}~\bibnamefont {Zoller}},\ }\href {\doibase
  10.1103/PhysRevLett.97.220403} {\bibfield  {journal} {\bibinfo  {journal}
  {Phys. Rev. Lett.}\ }\textbf {\bibinfo {volume} {97}},\ \bibinfo {pages}
  {220403} (\bibinfo {year} {2006})}\BibitemShut {NoStop}%
\bibitem [{\citenamefont {Chin}\ \emph {et~al.}(2010)\citenamefont {Chin},
  \citenamefont {Grimm}, \citenamefont {Julienne},\ and\ \citenamefont
  {Tiesinga}}]{RevModPhys.82.1225}%
  \BibitemOpen
  \bibfield  {author} {\bibinfo {author} {\bibfnamefont {C.}~\bibnamefont
  {Chin}}, \bibinfo {author} {\bibfnamefont {R.}~\bibnamefont {Grimm}},
  \bibinfo {author} {\bibfnamefont {P.}~\bibnamefont {Julienne}}, \ and\
  \bibinfo {author} {\bibfnamefont {E.}~\bibnamefont {Tiesinga}},\ }\href
  {\doibase 10.1103/RevModPhys.82.1225} {\bibfield  {journal} {\bibinfo
  {journal} {Rev. Mod. Phys.}\ }\textbf {\bibinfo {volume} {82}},\ \bibinfo
  {pages} {1225} (\bibinfo {year} {2010})}\BibitemShut {NoStop}%
\bibitem [{\citenamefont {Martikainen}(2011)}]{PhysRevA.83.013610}%
  \BibitemOpen
  \bibfield  {author} {\bibinfo {author} {\bibfnamefont {J.-P.}\ \bibnamefont
  {Martikainen}},\ }\href {\doibase 10.1103/PhysRevA.83.013610} {\bibfield
  {journal} {\bibinfo  {journal} {Phys. Rev. A}\ }\textbf {\bibinfo {volume}
  {83}},\ \bibinfo {pages} {013610} (\bibinfo {year} {2011})}\BibitemShut
  {NoStop}%
\bibitem [{\citenamefont {Paul}\ and\ \citenamefont
  {Tiesinga}(2013)}]{PhysRevA.88.033615}%
  \BibitemOpen
  \bibfield  {author} {\bibinfo {author} {\bibfnamefont {S.}~\bibnamefont
  {Paul}}\ and\ \bibinfo {author} {\bibfnamefont {E.}~\bibnamefont
  {Tiesinga}},\ }\href {\doibase 10.1103/PhysRevA.88.033615} {\bibfield
  {journal} {\bibinfo  {journal} {Phys. Rev. A}\ }\textbf {\bibinfo {volume}
  {88}},\ \bibinfo {pages} {033615} (\bibinfo {year} {2013})}\BibitemShut
  {NoStop}%
\bibitem [{\citenamefont {Isacsson}\ and\ \citenamefont
  {Girvin}(2005)}]{PhysRevA.72.053604}%
  \BibitemOpen
  \bibfield  {author} {\bibinfo {author} {\bibfnamefont {A.}~\bibnamefont
  {Isacsson}}\ and\ \bibinfo {author} {\bibfnamefont {S.~M.}\ \bibnamefont
  {Girvin}},\ }\href {\doibase 10.1103/PhysRevA.72.053604} {\bibfield
  {journal} {\bibinfo  {journal} {Phys. Rev. A}\ }\textbf {\bibinfo {volume}
  {72}},\ \bibinfo {pages} {053604} (\bibinfo {year} {2005})}\BibitemShut
  {NoStop}%
\bibitem [{\citenamefont {Collin}\ \emph {et~al.}(2010)\citenamefont {Collin},
  \citenamefont {Larson},\ and\ \citenamefont
  {Martikainen}}]{PhysRevA.81.023605}%
  \BibitemOpen
  \bibfield  {author} {\bibinfo {author} {\bibfnamefont {A.}~\bibnamefont
  {Collin}}, \bibinfo {author} {\bibfnamefont {J.}~\bibnamefont {Larson}}, \
  and\ \bibinfo {author} {\bibfnamefont {J.~P.}\ \bibnamefont {Martikainen}},\
  }\href {\doibase 10.1103/PhysRevA.81.023605} {\bibfield  {journal} {\bibinfo
  {journal} {Phys. Rev. A}\ }\textbf {\bibinfo {volume} {81}},\ \bibinfo
  {pages} {023605} (\bibinfo {year} {2010})}\BibitemShut {NoStop}%
\bibitem [{\citenamefont {Soltan-Panahi}\ \emph {et~al.}(2011)\citenamefont
  {Soltan-Panahi}, \citenamefont {L{\"u}hmann}, \citenamefont {Struck},
  \citenamefont {Windpassinger},\ and\ \citenamefont
  {Sengstock}}]{soltan2011quantum}%
  \BibitemOpen
  \bibfield  {author} {\bibinfo {author} {\bibfnamefont {P.}~\bibnamefont
  {Soltan-Panahi}}, \bibinfo {author} {\bibfnamefont {D.-S.}\ \bibnamefont
  {L{\"u}hmann}}, \bibinfo {author} {\bibfnamefont {J.}~\bibnamefont {Struck}},
  \bibinfo {author} {\bibfnamefont {P.}~\bibnamefont {Windpassinger}}, \ and\
  \bibinfo {author} {\bibfnamefont {K.}~\bibnamefont {Sengstock}},\ }\href@noop
  {} {\bibfield  {journal} {\bibinfo  {journal} {Nat. Phys.}\ }\textbf
  {\bibinfo {volume} {8}},\ \bibinfo {pages} {71} (\bibinfo {year}
  {2011})}\BibitemShut {NoStop}%
\end{thebibliography}%

\end{document}